\documentclass[aps,prd,twocolumn,groupedaddress,fleqn]{revtex4-2}
\usepackage{graphicx}
\usepackage{epsfig}
\usepackage{comment}

\usepackage{mathrsfs}
\usepackage{amssymb}
\usepackage[intlimits]{amsmath}

\begin{document}
\numberwithin{equation}{section}

\title{A conformal basis for cosmology with energy conservation}


\author{J. M. Greben}

\email[]{jmgreben@gmail.com}
\altaffiliation {Independent researcher since retiring as principal scientist from the Council of Scientific and Industrial Research (CSIR) South Africa}

\begin{abstract}
In the standard FRW formalism, the scale factor is assumed to describe the expansion of the universe. However, by examining empty space with a positive cosmological constant (i.e., a de Sitter space), we find that this assumption is incorrect. When described in conformal time, the associated conformal metric exhibits a big bang singularity where the effective vacuum energy density diverges, dominating the early universe.
The corresponding FRW scale factor decreases after the big bang, so that it does not describe the universe's expansion.  

Instead, the expansion is driven by global energy conservation: as the effective vacuum energy density decreases over time, space must expand to compensate. This leads to a linearly expanding universe, which agrees well with various cosmological observations - such as the red shifts of supernovae; the temperature dependence of the CMB radiation; and the near equality of the Hubble time and the age of the universe. Thus, the conformal vacuum metric can serve as a dominant background metric in cosmology, especially during the big bang era when it suggests new explanations for the creation of the initial energy and the first particles in the universe. 

Due to this dominance of vacuum energy, matter and radiation can be treated perturbatively. Even in their presence, the linear expansion remains the leading behavior, with the effective matter and radiation densities decreasing like $1/t^3$, just like the vacuum energy density. However, changes in their relative abundance can cause slight deviations from the linear expansion. Matter and radiation change the vacuum metric, thereby generating big secondary terms which could be responsible for the phenomenon of dark matter and for the acceleration of the expansion. 

\end{abstract}

\maketitle
\def\thesection{\arabic{section}}
\section{Introduction}
\label{sec:Introduction}
Given that dark energy is attributed to a (positive) cosmological constant $\Lambda$ in the $\Lambda$CDM model, one might ask whether the dominance of this vacuum energy - and the corresponding vacuum metric - could be extended back to the big bang. This would offer a more unified and cohesive picture of the universe’s evolution. However, in the standard FRW (Friedmann–Robertson–Walker) formalism \cite{Friedmann, Robertson, Walker} a vacuum-dominated universe results in an exponentially increasing scale factor. If this scale factor is taken to represent the cosmic expansion then it would imply an unphysical scenario: a continuous exponential expansion from the big bang onwards. Moreover, the exponential function lacks a finite point that could be associated with the big bang itself. 

However, this picture changes drastically if we use a metric that is conformal to the Minkowski metric and is expressed in conformal time. In this representation, the metric possesses a big bang singularity at a finite time where the effective vacuum energy density behaves like $1/t^3$. Thus, in conformal time, vacuum (or dark) energy plays a dominant role not just at late times but also during the big bang era. Since this picture contradicts the standard FRW picture, the question is: which description is physically correct, since the laws of physics should not depend on the choice of coordinates or metric?

To resolve this issue, we convert the conformal solution to the FRW metric and then find that it corresponds to the (usually ignored) FRW scale factor that decreases - rather than increases - after the big bang. In contrast to the ambiguity of the FRW solution — where either an increasing or decreasing exponential is possible — the conformal solution has a unique physically interpretation, being maximal at the big bang. So, the decreasing FRW solution must be chosen, implying that the FRW scale factor does not describe the expansion of the universe, as is usually assumed.

This result also implies that the evolution of the universe is best described in conformal time. Conformal time also provides a common language for cosmology, general relativity (GR) and Quantum Field Theory (QFT), as it is used in the description of quantum fields in curved space \cite{BD}. The conformal metric, being proportional to $1/t^2$, also features an elegant time symmetry. Hence, conformal time has many additional advantages. Despite the usage of conformal time and the important role of conformal transformations and the later introduction of a conformal scale factor, there is no direct link to other conformal theories like conformal gravity \cite{CG} and conformal field theory \cite{CFT}, as the current theory aims to stay within the usual boundaries of GR and QFT (except referring later to a bound-state extension of QFT \cite{QuarkSpatial}). 

Giving up on the FRW hypothesis - that its scale factor describes the expansion of the universe - is a big step, as it has dominated cosmology for a century without being seriously questioned. Hence, we comment briefly on possible shortcomings of its usual justification. Note first that its failure has come to light by analyzing the vacuum metric, whose importance for cosmology has long been doubted because of the smallness of the cosmological constant \cite{MTW, Carroll}. So traditionally the emphasis has been on idealized matter and radiation solutions, without allowing for the possibility that they could be perturbations on the dominant vacuum solution. The corresponding FRW scale factors are then required to vanish - rather than being maximal - at the big bang, as they are supposed to describe the expansion of the universe. 

The main justification for the FRW hypothesis is the assumption that the proper distance mirrors the global expansion of the universe, and that this proper distance is proportional to the FRW scale factor (see texts like \cite{MTW, Carroll, Weinberg}). However, the evidence for the expansion of the universe does not come from local measurements of the proper distance, but rather from astronomical measurements such as the red shifts of supernovae. These measurements refer to the far past, so they should preferably be interpreted by a theory that describes the global evolution of the universe over time. Apparently GR - being a local theory - cannot do this and we need a global theory or principle that can lay such a connection, presumably one with an integral (rather than a differential) character. 

Since global energy conservation can connect states at different times and can be expressed in integral form, it could play this global role. 
In QFT, global energy conservation can be derived from the condition that the divergence of the energy-momentum tensor vanishes. In GR its covariant generalisation in vacuum space can be converted into thermodynamic form, and then features a (negative) pressure term which is absent in QFT. It is this term which can explain the creation of vacuum energy in the beginning of the universe. This process owes its uniqueness to the fact that the negative pressure does not have a "physical" microscopic energy basis, but originates from the space-time structure of the vacuum universe. In 1990 Stenger \cite{Stenger} already demonstrated how this mechanism can create energy adiabatically, calling it the ultimate free lunch. Thus the absence of a time-like Killing vector in GR, and the consequential absence of a conserved energy current, serves a very fundamental function: namely to allow for the creation of the initial energy in the universe. 

Once this (vacuum) energy has been created, its density decreases rapidly because of the metric factor $1/t^3$ which multiplies the energy density $\epsilon=\Lambda/8\pi G$. Hence, the energy contained in a fixed volume decreases accordingly. Therefore, the volume must expand like $t^3$ to ensure the conservation of energy. As vacuum space is completely homogeneous, this implies a uniform expansion of space. Phrased differently: energy conservation, and not GR, causes the expansion of the universe. 

As energy conservation does not follow from GR, it must be imposed explicitly in cosmology. This has become possible now that the FRW scale factor no longer controls the expansion of the universe, so that the expansion can be controlled by energy conservation. Having a global character, the conservation condition does not interfere with the local divergence condition, which remains an important constraint on the metric. 

The expansion of the universe can now be modeled by a conformal scale factor which increases linearly in time. In Sec.\ \ref{sec:Matter} we show that it even keeps its linear character after the introduction of matter and radiation, except during periods when major changes occur in the composition of the universe.
The conformal scale factor thus takes over the role of the FRW scale factor, just like conformal time takes over the role of the FRW time variable. So, in the analysis of cosmological data, the FRW scale factor can simply be replaced by the conformal one. 
Note that the standard FRW scale factors (exponential for the vacuum universe, $\sim t^{2/3}$ for the matter and $\sim t^{1/2}$ for the radiation dominated universe) are not linear. Since the linear scale factor gives an excellent description of many cosmological data, this signifies an important advantage of the new formulation.

In Sec.\ \ref{sec:Red_shifts} we discuss a variety of cosmological data to demonstrate this good performance of the linear scale factor. First, the linearity implies that the Hubble time (the inverse of the Hubble constant) equals the age of the universe. The data confirm the (approximate) validity of this identity. In the $\Lambda$CDM model this equality can only be satisfied for a particular choice of the $\Lambda$CDM parameters (in particular that of $\Omega_m$), and thus would only hold at one particular time. The linear expansion factor also performs well in explaining the red shifts of supernovae. This was already demonstrated in 2011 using the then available data \cite{FOS_Greben}, an analysis which also provided a prediction for the age of the universe ($13.8\times10^9$ years) which agrees well with current estimates \cite{Bennett}. 

Since then, this good performance of linear scale factors has been confirmed in various other studies. For example, Traunmüller \cite{Traunmuller} fitted type 1A supernovae data without making any prior assumptions about the cosmological model and found that the best fit was obtained with a luminosity expression which is identical to the one obtained with our linear scale factor \cite{FOS_Greben}. Melia et al. \cite{Wei_Melia_2015} analyzed the Supernova Legacy Survey Sample \cite{Guy2010} and also found that the linear expansion factor does better than the $\Lambda$CDM model. 
The last observable considered in Sec.\! \ref{sec:Red_shifts} is the temperature dependence of the CMB radiation \cite{T_CMB}. Again, the linear scale factor gives a very good fit.

We start the discussion of the inclusion of matter by asking the question: how could (the first) particles be created in the empty vacuum universe without any prior particles being present? In Sec.\! \ref{sec:Early} we give a tentative answer to this question by linking the current theory to a
QFT theory which models elementary particles (quarks and leptons) as Planck-sized systems \cite{QuarkDressing, QuarkSpatial}. This theory suggests that when the vacuum energy density has decreased to the Planckian density $G^{-2}$ (about $10^{-24}$ seconds after the big bang), the circumstances are uniquely suited for this creation process as the positive vacuum energy density exactly matches the internal negative field energy of the particle. This theory also lays a connection between particle physics units and $G$ and $\Lambda$.  

In Sec.\! \ref{sec:Matter} we discuss the inclusion of matter and radiation in the conformal theory. In view of the assumed dominance of vacuum (dark) energy, we treat these components perturbatively. Because of the perturbative approach, the cosmological GR equations for matter and radiation can to first order be factorized into separate linear equations. The pointlike nature of gravitational objects in the universe can now be taken into account, enabling an important refinement of the cosmological principle. Except for periods when the relative percentages of the different constituents of the universe change globally, the linear expansion continues. This is an essential difference with the FRW formalism, where the three scale factors all behave differently with time, so that they determine the relative abundance of these components in different time periods. In the conformal formalism these ratios are no longer controlled by GR, but must be inferred from empirical observations and the astrophysical analysis of the transition processes which take place in different epochs. Changes in these abundances can then cause deviations from the dominant linear expansion.

The presence of matter and radiation modifies the vacuum metric, inducing secondary terms in the (global) energy integral. The mixed matter-vacuum energy term is a factor 9/4 larger than the original matter term, hinting at a possible cause of "dark" matter. In the radiation case there is an opposite effect as the factor equals -1/2, implying a reduction of the radiation energy. This has an interesting consequence: if there is a global nett conversion of matter into radiation in the late universe, than this will lead to an acceleration of the expansion, so that the induced energy loss can be compensated for.

Before starting the technical discussion, we comment briefly on the question: how can the $\Lambda$CDM model fit most cosmological data so well, when it relies to a considerable extent on the disputed FRW hypothesis? Part of the explanation may be that this model features many parameters and describes different periods using different models without much coherence between them, given it a lot of malleability. To illustrate this point, we recall that the linear scale factor gives a good description of a lot of cosmological data. If one tries to mimic this linear behavior with FRW scale factors, than these must be combined in a particular way. The matter ($\sim t^{2/3}$) and radiation scale factor ($\sim t^{1/2}$) do not suffice to reproduce this linear behavior, so the exponential scale factor (representing the dark energy in the $\Lambda$CDM model) needs to be introduced at some point. However, this exponential function will eventually surpass the linear trend, so that its contribution must decrease eventually. This phenomenon is indeed observed in the recent Dark Energy Survey \cite{DES}. 

In contrast, the conformal theory has a more cohesive character, being based on a continuous dominant vacuum background which starts already at the big bang. Also, it features only two basic parameters: $G$ and $\Lambda$. 

\section{Analysis of the FRW hypothesis}
\label{sec:Scale factor}
In the FRW formalism the time dependence of the cosmological metric is captured by the scale factor $a(t_{RW})$ which only multiplies the spatial part of the metric:
\begin{equation}
 \label{eq:FRW metric}
 ds^2=g_{\mu\nu}dx^{\mu}dx^{\nu}=-dt_{RW}^2+a^2(t_{RW})d\sigma^2.
\end{equation}
Although time is usually indicated by the symbol $t$ in this formalism, we reserve it for the conformal time, as this time variable plays a more central role in our formulation. 

As indicated originally by precision measurement of the CMB radiation \cite{Spergel} there is a lot of evidence that space is flat, with a recent survey \cite{Bennett} giving the value $\Omega_K=-0.0027\pm 0.0038$ for the curvature parameter. So we feel justified in setting:
 \begin{equation}
 \label{eq:spatial metric}
 d\sigma^2=dx^2+dy^2+dz^2,
\end{equation}
which also concurs with our desire to formulate cosmology as simple and elegant as possible.

The fact that the proper distance is proportional to $a(t_{RW})$ has given rise to the FRW hypothesis that the scale factor predicts the expansion of the universe (see in particular Ref.\! \cite{MTW}). 
In the introduction we gave various reasons why the validity of this FRW hypothesis should be investigated. 
We do this by comparing the physical properties of the de Sitter universe using different metrics, being mindful of the fact that the physical properties should not depend on the choice of the metric or the coordinate system. 

The energy-momentum tensor for the vacuum space with a positive cosmological constant has the form:
\begin{equation}
 \label{eq:Tuv_vacuum}
 T_{\mu\nu}(x)=-\epsilon g_{\mu\nu}(x),
\end{equation}
where the vacuum energy density $ \epsilon$ is related to the cosmological constant $\Lambda$ by the identity:
\begin{equation}
 \label{eq:epsilon}
\epsilon=\frac{\Lambda}{8\pi G}.
 \end{equation}
The two FRW solutions for this space are given by:
\begin{equation}
 \label{eq:a(t_{RW})}
 a(t_{RW})\sim\exp (\pm Ht_{RW})=\exp (\pm t_{RW}/t_H),
 \end{equation}
where 
\begin{equation}
 \label{eq:t_H}
H=\sqrt{\Lambda/3}=\sqrt{8\pi G\epsilon /3} \text{ and } t_H=\sqrt{3/8\pi G\epsilon}.
 \end{equation}
Here we use the symbol $H$, as this parameter would equal the Hubble constant if the + solution described the expansion of the universe. Its inverse $t_H=1/H$ would then stand for the Hubble time. 

Note that the FRW vacuum solution is not unique, as it can either be increasing or decreasing in time. If one assumes that the FRW scale factor describes the expansion of the universe, then one would choose the solution that increases with time. However, this is exactly the assumption we want to put to the test.

The second metric which we consider incorporates the time dependence in a conformal factor $\Omega(t)$, which multiplies the Minkowski metric used in QFT: 
\begin{equation}
 \label{eq:Conformal_metric}
 ds^2=\Omega^2(t)(-dt^2+d\sigma^2).
\end{equation}
Here $t$ is known as conformal time, sometimes indicated by the symbol $\eta$. 
GR yields two equations for $\Omega(t)$ (the derivatives are with respect to conformal time):
\begin{equation}
 \label{eq:Omega equations}
  \dot{\Omega}^2=\frac{1}{2}\Omega\ddot{\Omega}\text{  and   } \dot{\Omega}^2+\Omega\ddot{\Omega}=8\pi \epsilon G\Omega^4.
\end{equation}
These equations have the elegant solution:
\begin{equation}
 \label{eq:Omega solution}
  \Omega(t)= t_H/t,
\end{equation}					
where the constant $t_H$ was already defined in Eq.\ (\ref{eq:a(t_{RW})}). The singular point $t=0$ is now identified with the big bang. At this point the effective vacuum energy density (to be defined later) becomes infinite. Remarkably, this elegant, time-symmetric solution is rarely mentioned in the literature (an exception is the overview in \cite{Ibison}). Note that the same solution can be represented for $t\leq 0$, by changing the sign in Eq.\ (\ref{eq:Omega solution}), so as to keep $\Omega$ positive. 

Contrary to the FRW case, the two conformal solutions (for positive and negative $t$) describe the same physics, as in both cases the conformal factor is maximal near the big bang. The two solutions just refer to opposite sign conventions for time. Hence, there is no physical ambiguity in the conformal case. 

Since the conformal and FRW metric should yield the same physics, the physically unique conformal result can resolve the ambiguity in the FRW case. So let us construct the corresponding FRW solution. We have:
\begin{equation}
 \label{eq:Omega a}
  \Omega(t)=a(t_{RW}),
\end{equation}
and the differential condition:
\begin{equation}
 \label{eq:Omega a_condition}
  \Omega(t)dt= dt_{RW}.
\end{equation}
After inserting Eq.\ (\ref{eq:Omega solution}) in Eq.\ (\ref{eq:Omega a_condition}), we obtain the following relationship between $t$ and $t_{RW}$:
\begin{eqnarray}
\begin{aligned}
 \label{eq:t_nu}
&t_{RW}=t_H+t_H \ln{(t/t_H)} \leftrightarrow
\\
&t=t_H\exp[(t_{RW}-t_H)/t_H].
\end{aligned}
\end{eqnarray}
Here we have chosen the integration constant such that the proper time and the conformal time coincide at the special point $t=t_H$, when the conformal factor and (thus) the FRW scale factor equal unity.
The FRW scale factor $a(t_{RW})$ equals the conformal factor $\Omega(t)$, so that:
\begin{equation}
 \label{eq:FRW_a(tau)}
a(t_{RW})=t_H/t=\exp[ -(t_{RW}-t_H)/t_H].
\end{equation}
Hence, $a(t_{RW})$ is maximal at the big bang, when $t=0$ or $t_{RW}\downarrow -\infty$. This result would not have changed if one had chosen the opposite sign in Eq.\ (\ref{eq:Omega a_condition}), as it would simply leads to an inversion of the time direction with the big bang in the FRW formalism lying at $+\infty$. 

The ambiguity in Eq.\ (\ref{eq:a(t_{RW})}) is thus resolved, with the exponential that decreases after the big bang being the physically correct one. However, this also implies that the FRW scale factor does not describe the expansion of the universe, a drastic departure from the usual assumption in the FRW formalism. Hence, the expansion of the universe needs a new explanation. In the next section we will see that it is driven by the demand of energy conservation. 

\section{Energy conservation in vacuum space}
\label{sec:EC}
The standard derivation of energy conservation in QFT is based on the zero divergence of the energy-momentum tensor. So let us consider its consequences in GR, where the divergence is replaced by its covariant generalization:
\begin{equation}
 \label{eq:nabla_Tuv}
\nabla^{\mu}T_{\mu\nu}=0 \text{   or    } \nabla^{\mu}T_{\mu}^{\ \nu}=0 .
\end{equation}
This condition is automatically satisfied for the vacuum energy-momentum tensor, as $\nabla^{\mu}g_{\mu\nu}=0$ is a basic condition of GR. In analogy to QFT, we carry out a spatial integral over this divergence: 
\begin{eqnarray}
\begin{aligned}
 \label{eq:Energy_conservation}
&0=\int_V d^3x\sqrt{^3g}\ \nabla_{\mu}T^{\mu}_0=
\\
&=\int_V d^3x\sqrt{^3g}\left(\partial_{\mu}T^{\mu}_0-\Gamma^\lambda_{\mu 0}T^{\mu}_\lambda+\Gamma^\mu_{\mu \lambda}T^{\lambda}_0\right).
\end{aligned}
\end{eqnarray}
Here $^3g=\Omega^6(t)=a^6(t_{RW})$ is the spatial part of the metric determinant $|g|$, also called $|\gamma_{ij}|$ in some texts (see for example \cite{Weinberg}). 
This Jacobian factor is needed to ensure the invariance of the volume element in GR. In the standard 4D-integrals (such as in the action) one encounters the invariant volume element $d^4x\sqrt{|g|}$. However, for spatial integrals (which can easily be defined for a diagonal metric like the current one), the reduced determinant $^3g$ should be used. We will see that its 3D-character is essential for getting correct results. The Christoffel symbols for this vacuum space read:
 \begin{equation}
 \label{eq:Christoffel}
\Gamma^\lambda_{\mu \nu}=-\frac{1}{t}\left(\delta_{\mu 0}\delta_{\nu \lambda}+\delta_{\nu 0} \delta_{\mu \lambda}-\eta^{\lambda 0}\eta_{\mu \nu}\right),
\end{equation}
so that Eq.\ (\ref{eq:Energy_conservation}) can be written as:
\begin{eqnarray}
 \label{eq:Integral_Tuv}
0=\int_V d^3x\sqrt{^3g}\left(\partial_0 T^0_0-\frac{3}{t}T^0_0+\frac{1}{t}\sum_i T_i^{\, i}\right).
\end{eqnarray} 	
We now introduce the perfect fluid parametrization which is common in cosmology and exact for the vacuum universe:
\begin{equation}
 \label{eq:perfect_fluid}
T_{\mu}^{\, \nu}= -\delta_{\mu}^{\, \nu}(\delta_{\mu 0}\ \rho - \bar{\delta}_{\mu 0}\ p),
\end{equation}
where $\rho$ and $p$ are the energy and pressure density. For the de Sitter space the energy density $\rho$ is constant and equal to $\epsilon$, while the pressure density $p$ is negative and equal to $-\epsilon$. Hence:
\begin{eqnarray}
\begin{aligned}
 \label{eq:EC_consequence}
&0=-\int_V d^3x\  \left\{\partial_0 ( \sqrt{^3g}\ \epsilon)-\frac{3}{t}\sqrt{^3g}\ p\right\}
\\
&= -\frac{d}{dt}\  \left \{\int_V d^3x \sqrt{^3g}\ \epsilon +\int_V d^3x\sqrt{^3g}\  p\right\},
\end{aligned}
\end{eqnarray} 		
which is also correct if $V$ depends on time (as $\epsilon=-p$). To allow for this possibility, we have replaced the partial derivative by a full derivative, which extends the explicit time dependence generated by GR with the implicit time dependence $(\partial V/\partial t)(\partial /\partial V)$. We note that the clean separation in an energy and pressure term depends crucially on the use of the reduced spatial metric factor $\sqrt{^3g}$.
 
Eq.\ (\ref{eq:EC_consequence}) suggests that for this diagonal cosmological metric and energy tensor, the energy can be defined as a straightforward generalisation of the expression in QFT:
 \begin{equation}
 \label{eq:E_V}
E_V=-\int_V d^3 x\sqrt{^3g}\ T_0^{\, 0}.
\end{equation}
In QFT, where $g_{00}=-1$, this can (and usually is) written as an integral over $T_{00}$, as $-T_0^{\, 0}=T_{00}$ in that case. 

A possible criticism of this definition could be that $T_0^{\ 0}$ is a component of a tensor and "thus" not invariant under space-time transformations. This would be true if $T_{00}$ would be used in its definition. However, $T_0^{\ 0}=-g_0^{\ 0}\epsilon=-\delta_0^{\ 0}\epsilon= -\epsilon$ is constant and independent of the metric. 
Another objection is advanced by Carroll (\cite{Carroll}, p.\ 120), who states that this definition implies that the energy increases exponentially, which he finds an unacceptable violation of energy conservation. But this argument is based on the usual FRW hypothesis where the metric factor $\sqrt{^3g}=a^3(t_{RW})$ increases exponentially. Now that we have shown that  $a(t_{RW})$ decreases in time, the conservation of $E_V$ can be ensured by increasing $V$, as we will show below. 

Naturally, as soon as matter and radiation are included, these also become part of the energy integral. In their presence, space is no longer purely homogeneous, so that the volume $V$ must be large enough to maintain the underlying cosmological principle.

The expression for the energy in a volume $V$ in Eq.\ (\ref{eq:E_V}), suggests that the metric factor $\sqrt{^3g}$ can also be interpreted as a (time-dependent) modifier of the vacuum energy density. We thus define the effective vacuum energy density:
\begin{equation}
 \label{eq:epsilon_eff}
 \epsilon^{eff}(t)=\sqrt{^3g(t)}\ \epsilon=\frac{t_H^3}{t^3}\epsilon.
\end{equation}
The fact that this density approaches infinity near $t=0$ is the reason why it can be identified with the big bang and shows that vacuum energy and the cosmological constant dominate in this phase. It also shows that the metric factor, which has its origin in the definition of the proper volume, can also serve in another capacity in coordinate space. Such duality is quite common in physics, e.g., the cosmological constant was originally introduced as a free geometric parameter in GR, but now serves mainly as the basis for vacuum (dark) energy.

Continuing the discussion of the physical implications of the divergence relation, we introduce the notation $\hat{V}$ for the invariant volume using the original meaning of the metric factor:
\begin{equation}
 \label{eq:V_hat}
\hat{V}=\int_V d^3 x\sqrt{^3g}.
\end{equation}
We can then write  Eq.\ (\ref{eq:EC_consequence}) in the form of the first law of thermodynamics:
\begin{equation}
 \label{eq:first law}
dE_V + p \ d\hat{V}=0.
\end{equation}
There is no entropy term in this equation, which is consistent with the fact that the conformal vacuum solution is unique in the classical vacuum space. 

Stenger \cite{Stenger} derives a similar equation, using more qualitative arguments. He then reasons that the creation of vacuum energy $E_V$ can be understood as an adiabatic process, whereby the negative pressure $p$ does work on itself, thereby raising the free energy of the universe as it expands. He calls this the "ultimate free lunch": getting something for nothing. 

Although Stenger formulates his arguments in the usual FRW context, it can be adapted to the current framework. As noted after Eq.\ (\ref{eq:EC_consequence}), the full derivative $d/dt$ is the sum of explicit and implicit derivative terms. Only the implicit terms (the derivatives with respect to $V$) are associated with the expansion of the universe and play a role in the thermodynamic analysis. So we write:
\begin{equation} 
 \label{eq:first_law_real}
\partial_V E_V +p\ \partial_V \hat{V}=0.
\end{equation}   
Both $E_V$ and $\hat{V}$ will increase if $V$ increases. Following Stenger, we can now argue that the negative pressure $p$ does work on itself and raises the "free" energy $E_V$, while the universe expands. So this mechanism explains how vacuum energy can be created in the vacuum space.

Once produced, $E_V$ defines the energy of the system and must be conserved on its own. This matches the way energy conservation is formulated in QFT, where the pressure term is absent and the creation of (vacuum) energy would not be explicable.
Hence, the condition for energy conservation is now expressed as a full derivative:
\begin{equation}
 \label{eq:E_V_conservation}
dE_V =0.
\end{equation} 
Eq.\! (\ref{eq:first law}) and Eq.\! (\ref{eq:E_V_conservation}) then imply that $\hat{V}$ is not only invariant under coordinate transformations, but also in time: 
\begin{equation}
 \label{eq:V_conservation}
d\hat{V} =0.
\end{equation}
Writing out Eq.\ (\ref{eq:E_V_conservation}), we get:
\begin{equation}
 \label{eq:Differential_law}
\frac{dE_V}{dt} = \epsilon V \frac{\partial \sqrt{^3g} }{\partial t}  + \epsilon\sqrt{^3g}\ \frac{\partial V }{\partial t} =0.
\end{equation}
Hence:
\begin{equation}
 \label{eq:V_eta}
V \rightarrow V(t)=\frac{t^3}{t_H^3}V_H,
\end{equation}
as $\partial \sqrt{^3g}/\partial t =-(3/t)\sqrt{^3g}$. Here $V_H$ is the reference value of $V$ at $t=t_H$. So, the expansion of the universe is necessary to satisfy the global conservation of energy. 

The increase in volume can be seen as a uniform linear expansion, i.e.\! a scale transformation which only shrinks or stretches space without changing its geometry. We represent it by the conformal scale factor:
\begin{equation}
 \label{eq:Conf_scale_factor}
a^{con\!f}(t)=\frac{t}{t_H}.
\end{equation} 	
This expansion does not have a dynamical character as it is a response to a global condition and does not follow from the solution of local dynamical equations. The degree of expansion is part of the specification of the state of the universe and as such can be seen as an external input in relation to local dynamical theories. Therefore, the conformal scale factor should not be used inside local GR or QFT equations. However, it must be used in determining and analyzing entities which depend on the expansion of the universe, such as the Hubble constant and the red shifts of supernovae.

The conformal factor $\Omega(t)$ multiplied the full space-time Minkowski metric, preserving its space-time symmetry. So, it is natural to do the same for the conformal scale factor, turning it into a true scale transformation. We then can define the co-moving space-time variable $\tilde{x}^{\mu}$, which does not change under the expansion of the universe:
\begin{equation}
 \label{eq:co_moving}
\tilde{x}^{\mu}=\frac{x^{\mu}}{a^{conf}(t)}=\frac{t_H}{t} x^{\mu}\equiv \frac{t_H}{t} (t,\vec{x}).
\end{equation}
This shows that the co-moving time $\tilde{x}^0$ is constant: $\tilde{x}^0=t_H$. However, this does not mean that there is no dynamical variation in time, because:
\begin{equation}
 \label{eq:dynamic}
d\tilde{x}^0=d\left(\frac{t_H}{t}t\right)=\frac{t_H}{t}dt.
\end{equation}
as $t/t_H$, and thus its inverse, refers to the external scale factor $a^{conf}$, which is not part of the dynamics. This identity expresses formally what we explained before: that the implicit time dependence due to the expansion of the universe should not appear in dynamical equations, being an external scale factor.

Complementary to the co-moving spatial variable $\tilde{x}_{\mu}$, we can now define the co-moving 4-momentum $\tilde{p}_{\mu}$:
\begin{equation}
 \label{eq:comoving_momenta}
\tilde{p}_{\mu}= a^{conf}(t)p_{\mu}= \frac{t}{t_H}\ p_{\mu}\longleftrightarrow p_{\mu}=\frac{t_H}{t}\tilde{p}_{\mu}.
\end{equation}
This equation shows how the energy-momentum $p_{\mu}$ of a photon decreases with time due to the expansion of the universe (the red shift). This property is usually attributed to GR, but now is seen as a consequence of the physical expansion of the universe due to energy conservation. 
Note finally that the four-product $p\, .\, x=p_{\mu}x^{\mu}$, which plays an important role in QFT and is part of the plane wave description of propagating photons, is invariant under this scale transformation. So, the conformal scale factor and its linearity again demonstrate the close connection between the current conformal formulation and quantum physics. 

\section{The analysis of cosmological observables}
\label{sec:Red_shifts}
As we will see in Sec.\ \ref{sec:Matter}, even after the inclusion of matter and radiation, the linearity of the conformal scale factor persists, except for periods when the global composition of the universe changes considerably. So, it makes sense to analyze a range of cosmological observations using this linear scale factor, before discussing these secondary deviations. 

As a first test of this linearity, we consider the Hubble constant, whose value has been the subject of recent debates because of the so-called Hubble tension \cite{Tension, Valentino}. We find:
\begin{equation}
 \label{eq:Hubble_identity}
H(t)=\frac{\dot{a}^{conf}(t)}{a^{conf}(t)}=\frac{1}{t}.
\end{equation}
We thus see that the inverse of the Hubble constant (the Hubble time $H_0^{-1}$) equals the age of the universe:
\begin{equation}
 \label{eq:Hubble_time}
 H_0^{-1}=t_0.
\end{equation} 
Most recent analyses agree that the age of the universe equals about $13.8\times10^9$ years \cite{Planck, Bennett}. 
Eq.\! (\ref{eq:Hubble_time}) then implies that $H_0$ should equal about $70.9$ km/s/Mpc (in the usual Hubble units). This value is close to the value obtained in the WMAP mission: $69.3 \pm 0.8$ in \cite{Bennett}, and falls right in between the value $67.27\pm 0.60$ obtained by the Planck 2018 team \cite{Planck} and the value $73.04\pm 1.04$ obtained by the SHOES collaboration \cite{Shoes}. Hence, this identity is quite consistent with the current data.

None of the FRW scale factors can reproduce this identity, as $H_0t_0$ equals 2/3, 1/2 and $H t_0$, respectively, for the matter, radiation and vacuum universe ($H$ is the constant defined in Eq. (\ref{eq:a(t_{RW})})). 
As the actual FRW scale factor is a weighted mixture of these three components, Eq.\! (\ref{eq:Hubble_identity}) can only be satisfied in the FRW formalism if the composition of the universe is adjusted continuously, implying that GR controls the time development of the composition of the universe. We consider this an overestimation of the power of GR, and attribute this consequence to the faulty FRW hypothesis.

Next we discuss the red shifts of supernovae, which provide a good test of cosmological models.
The red shifts are now analyzed using the conformal scale factor. The red shift parameter $z$ is defined in terms of the lengthening of the wave length since the time of emission $t_1$:
\begin{equation}
 \label{eq:z} 
z=\frac{\lambda_{obs}-\lambda_1}{\lambda_1}=\frac{a^{con\!f}(t_0)}{a^{con\!f}(t_1)}-1=\frac{t_0}{t_1}-1,
\end{equation} 
Here $\lambda_1$ usually corresponds to some well-known spectral line. 
In the analysis of supernovae data the luminosity distance plays an important role. This involves the calculation of the distance between the source and observer at the moment of detection. It is given by:
\begin{equation}
 \label{eq:distance} 
d(t_1,t_0)=c \int_{t_1}^{t_0} dt\frac{a^{conf}(t_0)}{a^{conf}(t)}=c\ t_0 \ln\left(\frac{t_0}{t_1}\right).
\end{equation}
For clarity we showed the light velocity $c$, which is normally omitted (as we set $c=1$). In order to obtain the luminosity distance $d_L$ one has to multiply this $d$ with the factor $a^{conf}(t_0)/a^{conf}(t_1)$ (see \cite{Weinberg} or \cite{Copeland}). Expressing the result in terms of $z$, we have:
\begin{equation}
 \label{eq:d_L_definition} 
 d_L=c\ t_0(1+z)\ln(z+1).
\end{equation}
This expression was already given in an earlier paper of ours in 2010 \cite{FOS_Greben}, and has the simple property that the deceleration, jerk, and snap parameter (as defined by Visser \cite{Visser}) all vanish. In \cite{FOS_Greben} we showed that Eq.\ (\ref{eq:d_L_definition}) gave a very good description of the (then available) supernovae data, with $t_0$ set to $13.8\times10^9$ years. This value agrees perfectly with the age of the universe listed in recent data surveys \cite{Planck, Bennett}. 

Since this paper appeared, the good performance of the linear scale factor has been confirmed by various other authors. In 2014 Traunmüller \cite{Traunmuller} carried out a phenomenological analysis of 892 type 1A supernovae data with a range of functional expressions for the luminosity distance, without making prior assumptions about the underlying cosmological model. He found that the optimal expression for the luminosity distance is exactly the one shown in Eq.\! (\ref{eq:d_L_definition}).

Melia and Shevchuk \cite{Melia2012} obtain a linear scale factor in their $R_h=ct$ model.
They analyze the Supernova Legacy Survey \cite{Guy2010} and claim that the linear model is more likely to be correct by $90\%$ compared to the $\Lambda$CDM model \cite{Wei_Melia_2015}.
Recently, this analysis was extended to the JWST data \cite{Melia2023}, also showing excellent agreement. 
Note that Melia derives the linear expansion factor using different arguments than we do, like the Birkhoff theorem and Weyl's hypothesis \cite{Melia2012}. But, since they assume that the scale factor must be based on the FRW formalism, they are forced to set $w=-1/3$ in the cosmological equation of state $p=w\rho$, as this is the only value which gives a linear FRW scale factor. This shows how the FRW hypothesis can lead to inconsistencies, as the only accepted values for $w$ are 0 (matter), $1/3$ (radiation) and $-1$ (vacuum). 
Another important cosmological observable is the temperature dependence of the CBM background radiation.
In Ref.\ \cite{T_CMB} indirect measurements are given for this quantity. Since, the photon energy decreasea like $1/t$, this dependence should look like: 
\begin{equation}
 \label{eq:CMB} 
T_{CMB}(z)\sim (z+1) T_{CMB}(0).
\end{equation}
This is in perfect agreement with the data and another indication that he vacuum background metric dominates the universe.

Finally, we discuss how the main parameter in the conformal theory - the cosmological constant $\Lambda$ - can be determined empirically. 
It is well-known that the density of the universe $\rho$ is very close to the so-called critical density $\rho_c$, where the latter is defined as  
\begin{equation}
 \label{eq:critical} 
\rho_c=\frac{3H^2}{8\pi G}.
\end{equation}
The reason that $\rho_c$ is called critical is that in the FRW formalism it marks the boundary between an open and a closed universe \cite{Carroll, Copeland}. In the conformal formalism, there is no such criticality, as the expansion is linear regardless of the value of the density parameter $\epsilon$. However, we can use the fact that $\rho /\rho_c \approx 1.02\pm 0.02$ \cite{Spergel}, to identify the effective density in the vacuum universe with the density $\rho_c$, so that:
\begin{equation}
 \label{eq:epsilon2} 
\epsilon \frac{t_H^3}{t_0^3} \approx \frac{3H_0^2}{8\pi G } =\frac{3}{8\pi G t_0^2} \rightarrow t_H\approx t_0, 
\end{equation} 
as $t_H^2=3/8\pi G\epsilon$. Since we have determined $t_0$ from red shift data, this give an estimate of $t_H$, and thus of the cosmological constant $\Lambda$. Hence, setting $t_H=t_0= 13.8\times 10^9 \text{ years}$, also yields estimates of the other constants: $\Lambda = 6.69 \times 10^{-84} \text{ GeV}^2$ and $\epsilon = 3.97 \times 10^{-47}\text{ GeV}^4$. 

Setting $t_0$ equal to $t_H$ also means that the current expansion factor $a^{conf}(t_0)=a(t_0)=1$. 
But this is exactly the condition which is imposed on the FRW scale factor. However, in the conformal case the normalization is already fixed, as $a^{conf}(t)=t/t_H$, so the fact that $t_0=t_H$ looks purely accidental (is it?), just like the property that the density of the universe happens to be close to the critical density. This is a puzzle still remaining.
\section{Early Universe}
\label{sec:Early}
Before discussing the inclusion of matter and radiation in the conformal cosmology, we pose the fundamental question: when and how did real (as opposed to virtual) particles first appear in the universe? Phrased differently: when did particles become part of the state vector of the universe and the source term of the Einstein equations? It is well-known that virtual particle-anti-particle pairs (quantum fluctuations) contribute to many physical processes, like the anomalous magnetic moment of electrons. However, the creation of real particles has only been observed when other particles were - or radiation was - already present.
All other constituents of the universe which we know of - like atoms, molecules and stars - had to wait for the right circumstances before they could make their entrance in the universe. The same may well be true for elementary particles. The question is thus: how could particles be created in the vacuum universe when no prior particles were present? 

In Sec.\ \ref{sec:EC} we saw how vacuum energy could be created at the big bang thanks to the negative pressure term in the vacuum metric.
Maybe the extreme high energy density of the early vacuum universe offers a unique environment for enabling the creation of the first particles. This is exactly what is suggested by a particle theory in which this high vacuum energy density plays an essential role.
Since a thorough discussion of this issue lies outside the scope of this paper, we only give a short sketch of how this creation process might take place. 

In this particle theory QFT is used to describe elementary particles (quarks or leptons) as finite systems, bound absolutely within a spherical enclosure by the gauge fields it generates itself \cite {QuarkDressing,QuarkSpatial}. This - originally scale-free - theory acquires its Planckian scale ($l_P= G^{1/2}= 1.6\times10^{-35}\text{m}$) through its interaction with GR. The link with the current cosmological theory lies in the fact that the negative internal quantum field energy of Planckian density ($\sim G^{-2}$) must be countered by positive vacuum energy with the same density, so that overall the system can acquire a positive energy (mass). 

In the early universe the vacuum energy has decreased to the Planckian density at a time $t=t_c$, when:
\begin{equation}
 \label{eq:Planck_domain}
\epsilon \frac{t^3_H}{t_c^3}=G^{-2}\rightarrow t_c= \left(\frac{3}{8 \pi}\right)^{1/2}\left(\frac{G}{\epsilon}\right)^{1/6} \approx 10^{-24}\,\text{s}.
\end{equation}
Surprisingly, this creation time lies right in the particle physics domain, equalling about $5.3\times 10^{-24}$ seconds. Its inverse $\hbar/t_c$ then lies in the particle physics MeV energy domain, and - as is argued in the particle paper - can be related to the mass of the light elementary particles. In passing this explains the origin of the particle physics scale in terms of the two fundamental cosmological constants $G$ and $\epsilon$ (or $G$ and $\Lambda$). 

In the current universe this Planckian creation process is hidden from view as the Planck length is 20 orders of magnitude smaller than the proton radius. Hence, the only property one needs to describe in particle physics is how particles (or radiation) supply the excess mass energy in the creation process. But in the early universe this process can take place spontaneously, as the vacuum energy with the correct density is naturally present and can induce the creation of elementary particles without prior particles being present.


The scenario does not only suggest how the first elementary particles can be created, but also when this will happen. Before $t=t_c$ only vacuum energy contributes to the source term in the Einstein equations, so that the classical vacuum metric reigns supreme in this period. Since the classical metric can deal with the singular nature of the big bang, this scenario also obviates the need for replacing the continuous description of space-time by a more nebulous space-time foam. The finite size and finite creation time screens QFT from dangerous point singularities, while in QFT's bound-state realization in the finite particle theory they are absent anyway.

It is hard to be specific about the evolution of the universe directly after this creation phase without developing the theory in more detail. The vacuum energy consumed in the creation process can be converted into the kinetic energy of the created particle-anti-particle pairs. But it also may be compensated for by an instantaneous increase in vacuum space (inflation). Such an increase would be tiny in comparison to that in the usual inflation theories. Because of the complete homogeneity in the original vacuum space, there is also the question whether such an inflationary period is necessary in the current theory to explain the uniformity of the CMB radiation. Furthermore, the created particles will be in very close causal contact directly after their creation and may well relax to a state of thermal equilibrium, so that the tiny temperature variations in their distribution might be explainable as quantum fluctuations. 

Whatever the outcome of such further studies, the current discussion already shows how the singular nature of the vacuum metric near the big bang opens up entirely new possibilities for handling the early universe and the interplay between QFT and GR in the Planck domain.

\section{Matter and Radiation}
\label{sec:Matter}
The previous section has shown that the linear conformal scale factor gives a good description of a range of cosmological data. Hence, the assumption that the vacuum metric can serve as a basic background metric seems justified. This assumption is also underpinned by the current consensus that dark energy dominates the energy content of the universe. So it seems logical to treat matter and radiation as a perturbation on the dominant vacuum energy. Of course, such a perturbative treatment can only be applied to the broad structure of the universe: compact astronomical objects like stars and black holes clearly cannot be described this way. 
The first step is to modify the de Sitter metric: 
\begin{equation}
 \label{eq:modified_gmunu}
\tilde{g}_{\mu \nu}(x)=\frac{t^2_s}{t^2}\left[\eta_{\mu \nu}+h_{\mu \nu}(x)\right],
\end{equation}
where in lowest order $h_{\mu \nu}$ can be written as a sum of matter and radiation components.

The energy-momentum tensor which drives the Einstein equations, is modelled according to the ideal perfect fluid model  which assumes that the energy is distributed homogeneously over the universe in accordance with the cosmological principle. Vacuum energy satisfies this principle exactly, however, matter and radiation only approximately. We will see that the current approach also allows for a refinement of this principle in the case of the matter distribution. We write (see Eq.\! (8.50) in \cite{Carroll}):
 \begin{eqnarray}
\begin{aligned}
 \label{eq:T*}
 &T_{\mu}^{(m)\nu}=- \delta_{\mu}^{\, \nu} \delta_{\mu 0}\ \rho^{(m)},
 \\
 &T_{\mu}^{(r)\nu}=- \delta_{\mu}^{\, \nu} \left( \delta_{\mu 0}\ \rho^{(r)}-\bar{\delta}_{\mu 0}\ p^{(r)}\right) ,
\end{aligned}
\end{eqnarray}
where $p^{(r)}=\rho^{(r)}/3$, so that the trace $T_{\mu}^{(r)\mu}=0$. 

To first order, the divergence relation, Eq.\! (\ref{eq:nabla_Tuv}), can be imposed separately on the matter and radiation energy tensor. Using the vacuum metric, i.e.\! using the Christoffel symbols given in Eq.\! (\ref{eq:Christoffel}), one finds:
\begin{equation}
 \label{eq:explicit_t}
T_{0}^{(m)0}=-\rho^{(m)}\sim t^3;\ T_{0}^{(r)0}=- \rho^{(r)}\sim t^4.
\end{equation}

In the conformal formulation one must also factor in the implicit time-dependence caused by the expansion of the universe.
For matter, we can do this by taking into account the near pointlike distribution of matter in the universe \cite{FOS_Greben}: 
\begin{equation}
 \label{eq:rho_matter}
\rho^{(m)}(x)=\sum_i \frac{M_i}{\sqrt{^3g}}\delta^{(3)}(\vec{x}-\vec{x}_i).
\end{equation} 
As Weinberg noted (Eq.\! (5.2.13) in \cite{Weinberg}), such a spatial $\delta$-function must be accompanied by the inverse function of the spatial metric. The explicit time-dependence in Eq.\ (\ref{eq:rho_matter}) is now represented by the factor $1/\sqrt{^3g}\sim t^3$, as required by the divergence relation. The implicit time-dependence is carried by the spatial $\delta$-function. It behaves essentially like $1/V$ (integrate over $V$ and then divide by it), and is thus proportional to $t^{-3}$. Hence, the explicit and implicit time-dependence cancel each other, so that over time the density is constant, unless transitions change the relative composition of the universe.

Eq.\ (\ref{eq:rho_matter}) represents a refinement over the usual cosmological representation of the matter density, where the spatial distribution is averaged from the start. By delaying the averaging process, we can take into account the special nature of the distribution, without actually having to know the explicit positions $\{\vec{x}_i\}$. The averaging over space is now postponed until after the Einstein equations have been solved, so that important terms can be picked up which otherwise are neglected. 

The radiation density can also be represented by a combination of explicit and implicit time dependence:
\begin{equation} 
 \label{eq:rho_rad}
\rho^{(r)}(t)= \frac{1}{\sqrt{|g|}} \frac{1}{V}\sum\limits_{i\in V} p^{(i)}_0.
\end{equation} 
In this case, both the position and the time of arrival of the radiated photons are fixed by the emitter. This space-time information necessitates the introduction of the full space-time metric factor $\sqrt{|g|}$ instead of $\sqrt{^3g}$. Since, photons do not compactify, the spatial dependence is averaged over. This leads to a sum over the photon energies $p^{(i)}_0$ divided by the volume $V$. 

The explicit time dependence is now carried by the factor $1/\sqrt{|g|}=t^4/t^4_H$, as required by the GR divergence condition. The photon energies behave like $1/t$ because of the red-shift, while the volume $V$ behaves like $t^3$, both being a consequence of the expansion of the universe. Together they behave like $1/t^4$, ensuring the constancy of $\rho^{(r)}$ over time, just like was the case for the matter density. As a result, all effective densities behave like $1/t^3$, apart from deviations due to mutual transitions.

From Eq.\ (\ref{eq:explicit_t}) we find that the explicit time dependence of the lower component of the energy momentum tensor $T_{\mu\nu}^{(m)}$ and $T_{\mu\nu}^{(r)}$ behaves like $t$ and $t^2$, respectively.  
Using this knowledge, we can solve the linearized first-order Einstein equations for $h^{(m)}_{\mu\nu}$ and $h^{(r)}_{\mu\nu}$. 
After cancelling out the pure vacuum terms, both functions satisfy the same equation (see also \cite{EC_cosmology}): 
\begin{eqnarray}
 \label{eq:h_equation}
 \begin{aligned}
&\frac{2}{t}\left(\partial_{\mu}h_{\nu 0}+\partial_{\nu}h_{\mu 0}-\partial_0 h_{\mu\nu}\right)-\frac{4}{t}\eta_{\mu\nu}\partial^{\alpha}(h_{0 \alpha}-\frac{1}{2}t_{0 \alpha}\hat{h})
\\
&-\frac{6}{t^2}\eta_{\mu\nu}\,h_{00}+\partial_{\mu}\partial^{\lambda}h_{\nu\lambda}
+\partial_{\nu}\partial^{\lambda}h_{\mu\lambda}-\partial_{\nu}\partial_{\mu}\hat{h}+
\\
&+\eta_{\mu\nu}\left(\Box \hat{h}-\partial^\alpha\partial^\beta h_{\alpha\beta}\right)-\Box h_{\mu\nu}=16\pi G\,T_{\mu\nu}.
 \end{aligned}
\end{eqnarray}
Here $\hat{h}$ is defined by:
\begin{equation}
\label{eq:h_definition}
\hat{h}=\eta^{\alpha\beta}h_{\alpha\beta},
\end{equation}
and $\Box = \eta^{\alpha\beta}\partial_{\alpha}\partial_{\beta}$.
The mixed vacuum-matter and vacuum-radiation terms can be recognised by their explicit $1/t$ or $1/t^2$ time-dependence. Dropping these would bring us back to the usual perturbative (weak gravity) equations (see e.g.\! Ch.\! 18 in \cite{MTW} or Eq.\! (10.1.4) in \cite{Weinberg}). Hence, this allows us to profit from the methods and conventions which have been developed in that field.

Since $T_{\mu\nu}$ is diagonal, the same can be assumed for $h_{\mu\nu}$. Isotropy also implies that the $h_{ii}$ are independent of the index $i$, so we set:
\begin{equation}
\label{eq:h}
h_{ii}=h,\ \forall i\ \rightarrow \hat{h}=-h_{00}+3h.
\end{equation}
For ${\{\mu\nu\}}={\{00\}}, $Eq.\ (\ref{eq:h_equation}) reduces to:
\begin{equation}
\label{eq:00}
\frac{6}{t^2}h_{00}-\frac{6}{t}\partial_0 h+\partial_0\partial_0 (h-h_{00})-2\triangle h=16\pi G T_{00},
\end{equation}
while for ${\{\mu\nu\}}={\{ii\}}$ we obtain:
\begin{eqnarray}
\begin{aligned}
\label{eq:ii}
&-\frac{6}{t^2}h_{00}+\frac{4}{t}\partial_0 h+\frac{2}{t}\partial_0 h_{00}-2\partial_0\partial_0 h+
\\
&+\triangle (h-h_{00})+ \partial_i \partial_i (h_{00}-h)=16\pi G T_{ii},
\end{aligned}
\end{eqnarray}
for fixed $i=1,2 \text{ or }3$. The individual $\partial_i \partial_i$ term should be absent because of isotropy, so we must set $h_{00}=h$. 
So, both in the matter and in the radiation case we can simplify the equations by setting:
\begin{equation}
\label{eq:h_mu_nu}
h_{\mu\nu}(x)=\delta_{\mu\nu} h(x).
\end{equation}
This was the condition we missed in earlier work \cite{FOS_Greben, EC_cosmology}. 

In the matter case the time-dependent terms cancel if we set $h^{(m)}\sim t$. We then get:
\begin{equation}
\label{eq:delta_h_m}
\triangle h^{(m)}(x)=-8\pi G\ T_{00}^{(m)}(x)=-8\pi G\ \frac{t_H^2}{t^2}\rho^{(m)}(x),
\end{equation}
so that
\begin{eqnarray}
\begin{aligned}
 \label{eq:h_m}
&h^{(m)}(x)=2G \frac{t_H^2}{t^2}\int d^3 x'\ \frac{\rho^{(m)}(t,\vec{x}')}{|\vec{x}-\vec{x}'|}=
\\
=&2G\frac{t}{t_H}\sum\limits_{i } \frac{M_i}{|\vec{x}-\vec{x}_i|}\theta\left(c(t-t_i)-|\vec{x}-\vec{x}_i|\right),
\end{aligned}
\end{eqnarray}
where the Heavyside-function $\theta$ is needed to guarantee causality. Now that we have obtained the solution of the Einstein equation, we can introduce the implicit time-dependence (the correct order). To this end we express the equation in terms of the co-moving variables $\tilde{x}^{\mu}$: 
\begin{equation}
 \label{eq:h_m_reduced}
h^{(m)}(x)\!=\!2G\!\sum\limits_{i}M_i \frac{\theta\left[c(t_H-\tilde{t}_i)-|\vec{\tilde{x}}-\vec{\tilde{x}}_i|\right]}{|\vec{\tilde{x}}-\vec{\tilde{x}}_i|} \!,
\end{equation}
where $\tilde{t}_i=(t_H/t)t_i$ and $\vec{\tilde{x}}_i=(t_H/t)\vec{x}_i$, in accordance with Eq. (\ref{eq:co_moving}).
Note that by using co-moving coordinates we have arrived at the usual Newtonian expression.

In the radiation case we get consistency by setting $h^{(r)}\sim t^4$. We then get the simple identity:
\begin{eqnarray}
\label{eq:h_rad}
h^{(r)}=-\frac{8 G\pi}{9} t^2 T^{(r)}_{00}=-\frac{8 G\pi}{9} t^2_H\, \rho^{(r)},
\end{eqnarray}
where we used Eq.\! (\ref{eq:explicit_t}) in the second step. Using Eq.\! (\ref{eq:t_H}) we then get:
\begin{eqnarray}
  \label{eq:h_rad2}
h^{(r)}=-\frac{1}{3 \epsilon}\rho^{(r)}=-\frac{1}{3\epsilon}\frac{t^4}{t_H^4}\frac{1}{V}\sum_{i\in V}p^{(i)}_0.
\end{eqnarray}
Note that the gauge freedom which exists in the standard weak gravity theory \cite{Franklin} is absent here because of the link to the dominant vacuum energy.

Having solved the Einstein equations, we can now account for the implicit time-dependence and express the solution in co-moving variables:
\begin{eqnarray}
  \label{eq:h_rad3}
\left<h^{(r)}\right>=\!-\frac{1}{3\epsilon}\frac{1}{V_H}\sum_{i\in V_H}\tilde{p}^{(i)}_0\!=\!-\frac{1}{3\epsilon}\left<\rho^{(r)}\right>.
\end{eqnarray}
Here we have introduced the average notation $<\cdots>$ to indicate that the resulting expressions are constant as far as GR and the expansion is concerned. Possible variations, due to transitions between different components, must be modelled separately, and will be discussed at the end of this section.



We now include the matter and radiation energy in the energy integral, also allowing for the first-order modification of the metric factor $^3g\rightarrow\, ^3\tilde{g}$:
\begin{equation}
 \label{eq:E_V_inclusive}
E_V=\!\int_{\hat{V}(t)} d^3x\sqrt{^3\tilde{g}}\ \left[\epsilon +\rho^{(m)}(x)+\left<\rho^{(r)}\right>\right],
\end{equation}
where:
\begin{equation}
 \label{eq:Vs_hat}
\hat{V}(t)= (t^3/t_H^3)\hat{V}_H.
\end{equation}
The reference volume $\hat{V}_H$ equals $V_H$ in the vacuum universe, but must change in the presence of matter and radiation to conserve energy. This is because the redistribution between vacuum, matter and radiation energy, together with the modification of the vacuum metric factor, changes the integrand. Since the energy in the co-moving volume must stay equal to $E_V$, the volume $V_H$ must then be adjusted, so that Eq.\! (\ref{eq:E_V_inclusive}) can remain valid. Since all corrections are treated to first order, the change in $V_H$ must also treated to first order. 


Note that the size of the reference volume is no longer arbitrary as the universe is no longer perfectly homogeneous. Hence, $V_H$ must initially be chosen large enough, so that at this stage the cosmological principle is still justifiable and the net inflow (outflow) of radiation in (or out) the co-moving volume can be ignored.

In first order, the corrections in the modified spatial metric factor $\sqrt{^3\tilde{g}}$ can be written as a sum:
\begin{eqnarray}
\begin{aligned}
 \label{eq:metric_factor_mod}
\sqrt{^3\tilde{g}(x)}&=\frac{t_H^3}{t^3}\left(1\!+h^{(m)}(x)+\!\left<h^{(r)}\right>\right)^{3/2}
\\
&\approx \frac{t_H^3}{t^3}\left(\!1\!+\!\frac{3}{2}h^{(m)}(x)\!+\!\frac{3}{2}\!\left<h^{(r)}\right>\right),
\end{aligned}
\end{eqnarray}
so that to this order Eq.\ (\ref{eq:E_V_inclusive}) becomes:
\begin{eqnarray}
\begin{aligned}
 \label{eq:E_V_first_order}
E_V&=\!\int_{\hat{V}(t)} d^3x \frac{t^3_H}{t^3} \left(\epsilon\!+\left<\rho^{(m)}\right>+\!\left<\rho^{(r)}\!\right>\right)\!+
\\
+&\frac{3}{2}\int_{\hat{V}(t)} d^3x \frac{t^3_H}{t^3}\epsilon\left( h^{(m)}(x)+ \left<h^{(r)}\right>\right),
\end{aligned}
\end{eqnarray}
where $\rho^{(m)}(x)$ was replaced by its expectation value, as it is integrated over the whole volume. Using Eq.\! (\ref{eq:rho_matter}) this expectation value can also be written as a sum over the masses contained in $\hat{V}(t)$. 

To calculate the integral over $h^{(m)}(x)$ is more tricky, as it features a double integral over $\rho^{(m)}(x)$. This involves matter densities which existed in the past. If we assume that these densities have remained constant over time, then we get approximately (see also \cite{FOS_Greben}):
\begin{equation}
 \label{eq:h_mat_average}
 \left <h^{(m)}\right >\approx\frac{3}{2}\frac{<\rho^{(m)}>}{\epsilon}.
\end{equation}
After carrying out the integration, we then can write:
\begin{eqnarray}
\begin{aligned}
\label{eq:Components_energy}
&E_V= \epsilon V_H =\epsilon \hat{V}_H\times 
\\
\times\!&\left\{1\!+\frac{\left<\rho^{(m)}\right>}{\epsilon}\, [1\! +\! c^{(m)}\, ]\!
+\!\frac{\left<\rho^{(r)}\right>}{\epsilon}\, [1\! +\! c^{(r)}\, ]\right\},
\end{aligned}
\end{eqnarray}
where the constants $c^{(m)}\approx 9/4$ and $c^{(r)}=-1/2$ represent the induced contributions which arise from the influence of matter and radiation on
the vacuum metric.

We see that the induced matter term is a factor 9/4 larger than the original matter term, suggesting that it could be a global representation of dark matter. In favour of this idea is the fact that the pointlike $\delta^{(3)}(\vec{x}-\vec{x}_i)$ distribution of the matter density has been replaced by the more spread out $1/|\vec{x}-\vec{x}_i|$ distribution of the induced matter term $h^{(m)}(x)$ (see Eq.\! (\ref{eq:h_m})). Such a distribution is not unlike the dark matter distribution around galaxies. What speaks against this interpretation is that the factor 9/4 is considerably smaller than the ratio 5.3 between baryonic and dark matter, as given in a recent survey \cite{Planck}. However, this large number could have been influenced by the adherence to the FRW hypothesis in these data analyses. 
Furthermore, the factor 9/4 may increase if the percentage matter was larger in the past (which we suspect to be the case). Another objection to the dark matter interpretation is that the induced term owes its big size to the integration over all of space, so that its localization seems insufficient to describe the dark matter halo near galaxies. But, the localisation may improve if the corrections are calculated to higher order. So, further study is necessary to clarify this issue.     

The energy expression in Eq.\! (\ref{eq:Components_energy}) also shows that the (linear) expansion is modified once the percentages of matter and radiation change.
We can express this through a function $r(t)$, which multiplies the original linear scale factor and equals unity before matter and radiation are present. 
Hence, the modified conformal scale factor is then written as:
\begin{equation}
 \label{eq:scale_factor_modified}
\hat{a}^{con\!f}(t)=r(t) a^{con\!f}(t) =r(t)\frac{t}{t_H},
\end{equation}
with $r(t)$ given implicitly by the identity:
\begin{eqnarray}
\begin{aligned}
 \label{eq:r(t)}
&r^{-3}(t)=V_H/\hat{V}_H=
\\
=1+&\, [1\! +\! c^{(m)}\, ]\frac{\left<\rho^{(m)}(t)\right>}{\epsilon} 
\!+\, [1\! +\! c^{(r)}\, ]\frac{\left<\rho^{(r)}(t)\right>}{\epsilon}.
\end{aligned}
\end{eqnarray}
By inserting the current estimates for the matter and radiation percentages (linked to the $\Omega_i$-parameters of standard cosmology),
we get an estimate of the current proportionality constant $r(t_0)/t_H$.

One might expect that over time, more and more (massive) astronomical bodies will decay and convert (part of) their mass into radiation. 
To indicate qualitatively how $r(t)$ would behave under such an evolution, we assume that the matter density decreases exponentially:
\begin{equation}
 \label{eq:rho_m_t}
<\rho^{(m)}(t)>=<\rho^{(m)}(t_o)>\text{e}^{-\kappa(t-t_0)}.
\end{equation}
We furthermore observe that in a quantum process energy must locally be conserved, so that:
\begin{equation}
 \label{eq:m_rad_cons}
\frac{d}{dt}\left[<\rho^{(m)}(t)>\!+\!<\rho^{(r)}(t)>\right]=0.
\end{equation}
Under these model assumptions one can find a closed solution for $r(t)$.
After some algebra we find:
\begin{eqnarray}
\begin{aligned}
 \label{eq:r(t)_formula}
&r(t)=\left[1+\, \left(1\! +\! c^{(r)}\, \right)\frac{\left<\rho^{(m)}(t_o)+\rho^{(r)}(t_o)\right>}{\epsilon}+\right.
\\
&\left.+\left(c^{(m)}\!-\!c^{(r)}\right)\frac{\left<\rho^{(m)}(t_o)\right>}{\epsilon}\text{e}^{-\kappa(t-t_0)}\right]^{-1/3}.
\end{aligned}
\end{eqnarray}

Since we are mostly interested in the question whether the expansion is accelerating and how fast, we also
show the result for time derivative of $r(t)$:
\begin{equation}
 \label{eq:dr(t)}
\dot{r}(t)=\frac{1}{3}\kappa\left(c^{(m)}-c^{(r)}\right)r^4(t)\frac{\left<\rho^{(m)}(t)\right>}{\epsilon}
\end{equation}
Hence, the conversion of matter into radiation leads to an acceleration of the expansion ($\dot{r}(t)>0$), which eventually dies out, to be replaced by the usual linear expansion. This might be the new explanation for the alleged acceleration of the expansion, which in the $\Lambda$CDM model is attributed to the cosmological constant.

\section{Summary}
In this paper, we propose a new cohesive cosmological framework, in which the vacuum de Sitter metric (expressed in conformal time) serves as a continuous background metric, with the corresponding vacuum energy dominating the energy content of the universe from the big bang onwards. Global energy conservation is added as an underlying principle in cosmology and is shown to govern the expansion of the universe.

This description becomes feasible after we found that a key assumption in standard cosmology - that the FRW scale factor describes the expansion of the universe - is untenable. In the vacuum universe two FRW solutions are possible: an exponentially increasing and an exponentially decreasing one. However,  if one uses the conformal metric one finds a unique solution. This solution is maximal near the big bang and matches the decreasing FRW scale factor. So, the FRW scale factor does not describe the expansion of the universe. 

Global energy conservation can now be imposed in cosmology, as the amount of vacuum energy can  be adjusted to satisfy it. This adjustment can be done through an expansion (or contraction) of space, as this action is no longer constrained by the FRW scale factor. 

The FRW scale factor now takes on a different physical role, namely to multiply the vacuum energy density with a time dependent factor $1/t^3$. To compensate this decrease, and to satisfy energy conservation, the (vacuum) universe must expand linearly. This expansion can be modelled by a linear (conformal) scale factor whose time dependence has an implicit character, as it does not interfere with - or enter explicitly in - the local equations of general relativity. Instead, it defines the degree of expansion of the universe and is part of the time-dependent definition of the state of the universe. 

Since the conformal - and not the FRW - scale factor - describes the expansion of the universe, it replaces the FRW scale factor in the interpretation of astronomical observations. As shown in this and other studies, a linear scale factor gives excellent fits to the observed supernova red shifts and other cosmological data. 

In the big bang era the dominance of vacuum energy and the vacuum metric is absolute due to the singularity at $t=0$. The creation of the initial vacuum energy can now be attributed to the negative pressure in the vacuum universe, as this acts as a free source of energy. This thermodynamic mechanism has been proposed before, however, it integrates naturally in the current description of the initial universe. 

This early universe, with its high vacuum energy density, may well offer the right circumstances for creating the first elementary particles without prior particles being present. Basing ourselves on a particle theory in which quarks and leptons are described as finite Planck-sized systems, we can even provide a quantitative estimate of when (after about $10^{-24}$ seconds) and how (by converting vacuum energy) this might take place. 
Clearly, further consequences of this scenario must be analyzed before we can judge its overall viability. 


In line with the assumed dominance of vacuum energy, matter and radiation are treated as perturbative corrections to the vacuum metric in the later universe. Their effect can then be described through linearized GR equations. The new formulation also allows for a refinement of the cosmological principle by formally accounting for the point-like distribution of matter in the universe. This extension could be a first step towards the incorporation of local mass concentrations (or voids) in global cosmological descriptions. 

The linearity of the cosmic expansion remains intact in the presence of matter and radiation as long as the composition of the universe remains constant. In transitional periods - such as when there is a net global conversion between matter and radiation - deviations from this linear behavior can arise. An empirical and astrophysical understanding of these transition processes is thus necessary to explain and describe these small deviations from the linear expansion in different epochs. We present a simple model to better understand these effects.

Through its effect on the vacuum metric, matter and radiation lead to big secondary terms in the overall energy, amounting to a factor +9/4 in the matter case and a factor -1/2 in the radiation case. The induced matter term may represent a global average over more localized effects, which act like dark matter. This possibility could be further investigated by taking into account higher-order terms to accentuate its local character. The two induced terms together might cause the observed acceleration of the cosmic expansion.
 
In summary, this conformal approach offers promising new perspectives on key cosmological questions. The predictions we have presented are consistent with observational data, though we recognize that the standard $\Lambda$CDM model has been tested against a broader array of phenomena and has achieved reasonable success, albeit with growing internal tensions. This success of the $\Lambda$CDM model, despite relying on the contested FRW assumption, might stem from its flexibility in using different models and assumptions in different epochs and its large number of adjustable parameters. In contrast, the conformal theory is markedly more economic and coherent, relying only on two fundamental parameters: the gravitational and the cosmological constant. 
\newline

\end{document}